\documentclass[utf8]{frontiersSCNS}
\usepackage{url,hyperref,lineno,microtype,subcaption}
\usepackage[onehalfspacing]{setspace}
\usepackage[utf8]{inputenc}
\usepackage{float}
\usepackage{tabularx}
\usepackage{flushend}
\usepackage{color}
\usepackage{afterpage}
\usepackage{siunitx}
\usepackage{longtable}

\def\keyFont{\fontsize{8}{11}\helveticabold}
\def\firstAuthorLast{Młyńczak {et~al.}}
\def\Authors{Marcel Młyńczak\,$^{1,*}$ and Hubert Krysztofiak\,$^{2}$}

\def\Address{
$^{1}$Institute of Metrology and Biomedical Engineering, Faculty of Mechatronics, Warsaw University of Technology, Warsaw, Poland \\
$^{2}$Department of Applied Physiology, Mossakowski Medical Research Centre, \\Polish Academy of Sciences, Warsaw, Poland}

\def\corrAuthor{Marcel Młyńczak, PhD\\8 Sw. A. Boboli Street, 02-525 Warsaw}
\def\corrEmail{marcel.mlynczak@pw.edu.pl}

\begin{document}
\onecolumn
\firstpage{1}

\title[Discovery of causal paths ...]{Discovery of causal paths in cardiorespiratory parameters:\\a time-independent approach in elite athletes}

\author[\firstAuthorLast ]{\Authors}
\address{}
\correspondance{}

\extraAuth{}

\maketitle

\begin{abstract}

Training of elite athletes requires regular physiological and medical monitoring to plan the schedule, intensity and volume of training, and subsequent recovery. In sports medicine, ECG-based analyses are well established. However, they rarely consider the correspondence of respiratory and cardiac activity. Given such mutual influence, we hypothesize that athlete monitoring might be developed with causal inference and that detailed, time-related techniques should be preceded by a more general, time-independent approach that considers the whole group of participants and parameters describing whole signals. The aim of this study was to discover general causal paths among cardiac and respiratory variables in elite athletes in two body positions (supine and standing), at rest. ECG and impedance pneumography signals were obtained from 100 elite athletes. The mean heart rate, the root-mean-square difference of successive R–R intervals (RMSSD), its natural logarithm (lnRMSSD), the mean respiratory rate (RR), the breathing activity coefficients, and the resulting breathing regularity (BR) were estimated. Several causal discovery frameworks were applied, comprising Generalized Correlations (GC), Causal Additive Modeling (CAM), Fast Greedy Equivalence Search (FGES), Greedy Fast Causal Inference (GFCI), and two score-based Bayesian network learning algorithms: Hill-Climbing (HC) and Tabu Search. The discovery of cardiorespiratory paths appears ambiguous. The main, still mild, rules best supported by data are: for supine - tidal volume causes heart activity variation, which causes average heart activity, which causes respiratory timing; and for standing - normalized respiratory activity variation causes average heart activity. The presented approach allows data-driven and time-independent analysis of elite athletes as a particular population, without considering prior knowledge. However, the results seem to be consistent with the medical background. Causality inference is an interesting mathematical approach to the analysis of biological responses, which are complex. One can use it to profile athletes and plan appropriate training. In the next step, we plan to expand the study using time-related causality analyses.

\tiny
 \keyFont{ \section{Keywords:} Athlete training adaptation biomarker, Cardiac function, Tidal volume, Cardiorespiratory causality, Elite athletes}
\end{abstract}

\section{Introduction}

\noindent Elite athletes require regular physiological and medical evaluation and monitoring for proper planning of the schedule, intensity, and volume of training (\cite{Meeusen2013}). Therefore, exercise scientists and sports physicians seek convenient biomarkers to evaluate the state of an athlete’s body during training to monitor homeostasis, maximize effect, and avoid over-training (\cite{Wiewelhove2015}).

Some of the most commonly used parameters are related to cardiac function. Heart rate monitoring is popular in sport and recreational activity, and widely used thanks to easy access to sophisticated tools, enabling beat-by-beat registration of electrocardiographic (ECG) signals and evaluation of heart rate variability (HRV) (\cite{Buchheit2014,Schmitt2015,Duking2016,Giles2016,Bellenger2016,Plews2017}).

Methodical acquisition of RR intervals, performed during different training periods, provides a chance to discern the proper course of HRV changes under the influence of exercise training, and possibly to recognize anomalous patterns indicating poor post-exercise recovery, sustained fatigue, impaired adaptation, and development of over-training syndrome.

Despite wide access, practical application of HRV parameters in sports training monitoring remains limited. The seemingly simple phenomenon, related to autonomic nervous system activity, defies simple evaluation, because of many modifying factors. An additional problem is the selection of optimal HRV parameters. There is no clear consensus as to which are best in training response evaluation. Is a single parameter enough, or will a set be more effective? Should more advanced mathematical methods be used for optimal modeling? There is growing interest in this field and recent studies have identified new directions (\cite{Sala2016,Sala2017}). 

Because there is a complex relationship between heart rate and breathing, taking breathing activity into account seems relevant and necessary for proper analysis. (\cite{Grossmann2007,Gasior2016,Sobiech2017}). The influence of inspiration and expiration is usually apparent in resting ECG as sinus respiratory arrhythmia (\cite{Larsen2010, Shaffer2014,McCraty2015}). Bidirectional neural relationships between cardiac functioning and the respiratory processes have previously been presented. The cardiorespiratory coupling effect, where heartbeats seem to coincide with specific respiratory phases, has been tested recently (\cite{Penzel2016,Sobiech2017}). The possible physiological mechanisms behind it appear to include increased sympathetic nervous activity, as well as changes in arterial blood pressure. The effect of baroreflex, with baroreceptors playing a crucial role in the adjustment of neural responses, was also extensively described (\cite{Reyes2013}).

The studies present concerns about whether respiratory control should be conducted in the HRV analysis. For example, \cite{Saboul2013} found that the RMSSD index is uninfluenced by respiratory patterns, for both spontaneous and controlled breathing. Nevertheless, the relations are more evident when different mathematical analyses are performed.

Therefore, the identification of an optimal mathematical method has an essential role in the daily practice. Usually, in sports science research methodology, a parameter's value is a dependent variable of exercise stimuli (top-down approach). However, in practical applications, the strategy can be reversed, with the optimal parameter as the one which best indicates and describes adaptation status (bottom-up approach). The relation between perception and performance outcomes can be "correlated" with the results of the objective analysis. For instance, an over-training syndrome is a major issue that negatively affects an athlete's performance, but it is not always subjectively felt in the same way. 

In that context, one should consider evaluating cross-dependencies or even causalities in recorded signals and calculated parameters. Causality is then the way to describe the relationships, also specifying the direction and the structure. Causal relations can be established as a directed acyclic graph (DAG), the type of Bayesian network, in which each node can represent a parameter, and each directed link has a probability measure. Also, the graph may be rewritten into structural equation models (SEM) to get coefficients, which express the strength of connections. For such DAGs, the do-calculus rules theorem was introduced to analyze the effects of interventions (\cite{Pearl1995}).

Other frameworks to analyze different types of causalities in physiological signals have been also proposed (\cite{Muller2016,Penzel2016,Javorka2016}). Nonlinear approaches, assuming cardiorespiratory interactions, were developed (\cite{Jamsek2004,Lopes2011}). Relatively newly recognized phenomena - phase synchronization between heartbeats and ventilatory signal, inverse respiratory sinus arrhythmia - have also been used (\cite{Bartsch2015,Kuhnhold2017,Mazzucco2017}). Information domain applications for more than three signals were presented (\cite{Wejer2017}). Granger-based causality was employed (\cite{Porta2017}), and also tested along with coherence measure and cross-sample entropy (\cite{Radovanovic2018}). Cardiorespiratory coordination, proposed by \cite{Moser1995}, defines the mutual influence of the onsets of cardiac and respiratory cycles on each other. Various methods were proposed to analyze this phenomenon (\cite{Riedl2014,Sobiech2017,Valenza2018}).

We hypothesize, that the main athletic rationale for causal path discovery is to profile athletes within a new causal domain, plan training modifications - considered to be interventions done to found cause variables (\cite{Pearl2010}), and track changes in objective cardiorespiratory responses. However, the methods mentioned in the previous paragraph are specifically intended to describe the systems' mutual temporal activity, not to propose directly the possible changes to training and causal-related parameters. We think that such approaches should be preceded by a more general one, which takes into account the parameters describing the entire segment of data, and tries to search for the structure of directional relationships.

Therefore, in this paper, we seek general time-independent causal paths between basic cardiac and respiratory variables in certain population, elite athletes, in two body positions (supine and standing) at rest. 

\section{Materials and Methods}

\subsection{Subjects and Device}

\noindent A group of 116 elite athletes (38 female; ages $24.4 \pm 6.3$) participated. Due to artifacts in signals gathered during examination, resulting from body movement and imprecise electrode mounting, data from 16 athletes could not be reasonably analyzed and were therefore excluded. The final study group consisted of 100 athletes (32 female; ages $24.6 \pm 6.4$). 

The study was carried out at the National Centre for Sports Medicine in Warsaw during the routine periodic health evaluation and medical monitoring program, 3-4 months before the 2016 Olympic Games in Rio de Janeiro. The study group comprised athletes in sports of differing type and intensity. Data on sex, height, and body mass in specific sports are presented in Table \ref{tab1}.

The study, including the consent procedure, was approved by the Ethics Committee of Warsaw Medical University (permission AKBE/74/17). All participants were informed about the general aim of the measurements, though not about the importance of breathing activity (\cite{Mortola2016}). Each subject had previously signed a consent form for the routine medical monitoring, which includes a statement of acceptance of the use of the results for scientific purposes.

Pneumonitor 2 was used to collect single-lead ECG signals (Lead 2), along with impedance pneumography (IP), which is related to respiratory activity (\cite{Mlynczak2017}). The IP signal was measured using the tetrapolar method, with the specified electrode configuration (\cite{Seppa2013}). Receiving electrodes were placed on the mid-axillary line at about 5th-rib level. Application electrodes were positioned on the same level on the insides of the arms. Standard Holter-type, disposable ECG electrodes were used. The sampling frequency was 250 Hz, sufficient in terms of heart rate variability analysis and over-sampled from a respiratory perspective (\cite{Task1996}).

\subsection{Protocol and Preprocessing}

\noindent The measurements were performed in a diagnostic room designated for cardiological examinations. Since the periodic health evaluation and medical monitoring are performed frequently for Olympic-level athletes, the diagnostic room and the measurement procedure (very similar to that for resting ECG) were familiar to them.

Each athlete was asked to lie down on the diagnostic (ECG) couch. After attachment of the electrodes and passage of a 10-minute stabilization phase, they were asked to remain supine and breathe freely (spontaneous breathing), and recording began. After 6 minutes, the athlete was asked to stand and again breathe freely while standing for another 6 minutes (\cite{Gilder2008,Sala2016}). The duration of analysis (about 6 minutes for each body position) seems appropriate for characterizing cardiac and respiratory activities in a "single" measurement. This is consistent with the method used in other studies, with HRV measurement in the supine position followed by the standing position (\cite{Gilder2008,Sala2016}). While this resembles the orthostatic maneuver, we took the latter only as an inspiration, performing the analysis for the entire supine and standing periods, i.e., without consideration of adaptation, recovery, etc. Physical data (height, body mass, and sex) were registered during the routine medical examination performed the same day. 

Pre-processing of the obtained ECG signal consisted of non-linear detrending for baseline alignment and finding R peaks based on the Pan-Tompkins algorithm. Raw IP signals were pre-processed by removing the cardiac component from the IP signal by subtracting the noise component derived from least mean square adaptive filtration, then smoothing with a 400 ms averaging window (\cite{Mlynczak2017b}). Then, we detected and delimited breathing phases by applying an adaptive algorithm to the differentiated, flow-related signal. We did not carry out the calibration procedure to transform impedance values into volumes, instead assuming that impedance changes had reproduced the tidal volume signal in terms of shape since linear fitting provides the best agreement between IP and the reference, pneumotachometry (PNT) (\cite{Mlynczak2015}).

We finally considered 10 cardiac and respiratory parameters, estimated for each participant, for the entire recording, separately for supine and standing:

\begin{itemize}
	\item mean heart rate (HR);
	\item root-mean-square difference of successive R–R intervals (RMSSD);
	\item natural logarithm thereof (lnRMSSD);
	\item mean respiratory rate (RR);
	\item ciRR - coefficient of variation of instantaneous breathing rate (iRR, calculated between inspiratory onsets);
	\item cInsT - coefficient of variation of the durations of the inspiratory phases (InsT);
	\item cExpT - coefficient of variation of the durations of the expiratory phases (ExpT);
	\item cInsV - coefficient of variation of the amplitudes of the inspiratory phases (InsV);
	\item cExpV - coefficient of variation of the amplitudes of the expiratory phases (ExpV); and
	\item breathing regularity (BR), as described in formula (1) - tanh operations are added to ensure that a range of $0\%$ - $100\%$ is preserved).
\end{itemize}

\begin{equation}
	BR\quad =\quad \left( 100\quad -\quad 20 \cdot \left( \tanh { \frac { { \sigma  }_{ iRR } }{ \overline { iRR }  } + } \tanh { \frac { { \sigma  }_{ InsT } }{ \overline { InsT }  } + } \tanh { \frac { { \sigma  }_{ ExpT } }{ \overline { ExpT }  } + } \tanh { \frac { { \sigma  }_{ InsV } }{ \overline { InsV }  } + } \tanh { \frac { { \sigma  }_{ ExpV } }{ \overline { ExpV }  }  }  \right)  \right) [\%]
\end{equation}

The differences between body positions were assessed using the paired T or Wilcoxon rank tests (depending on the normality of the parameters, checked using the Shapiro-Wilk test; all with a significance level of $\alpha = 0.05$). Signal processing was performed using MATLAB software. Graphics and statistical inference were obtained using R software (\cite{R}). The dataset of parameters and the R script will be provided as a supplement to this paper to ensure reproducibility.

\subsection{Time-independent causal path discovery}

\noindent We started with the assumption that the general time-independent causality can be revealed only when the correlation appears meaningful. Therefore, we calculated the Bayesian correlation coefficient as a result of the multiplication of the slope coefficient from the linear model and the ratio of standard deviations of both vectors. Assuming

\begin{equation}
	X\quad =\quad \alpha \quad +\quad \beta Y\quad +\quad \varepsilon 
\end{equation}

\noindent estimated using Bayesian approach, then

\begin{equation}
	cor(X,Y)\quad =\quad \hat { \beta  } \cdot \frac { \sigma (Y) }{ \sigma (X) }  
\end{equation}

The significance was assumed when the maximum probability of effect $MPE > 0.9$ (from the estimation of the linear model) (\cite{Makowski2018}).

Then, we studied all pairs using 6 techniques:

\begin{itemize}
	\item the generalized correlations $ { r }_{ x|y }^{ * } $ and $ { r }_{ y|x }^{ * } $, with $ |{ r }_{ x|y }^{ * }| > |{ r }_{ y|x }^{ * }| $ suggesting that y is more likely to be the "kernel cause" of x (though only when the p-value is significant) - equation (4) below implements the generalized correlation (\cite{Vinod2017}) using the \textit{generalCorr} R package (\cite{generalCorr});
	\item causal additive modeling, with \textit{selGAM} pruning (\cite{Buhlmann2014}), using the \textit{CAM} R package (\cite{CAM});
	\item Fast Greedy Equivalence Search for continuous variables (\cite{Ramsey2015}) using the \textit{rcausal} R package (\cite{rcausal});
	\item Greedy Fast Causal Inference for continuous variables (\cite{Ogarrio2016}) using the \textit{rcausal} R package (\cite{rcausal});
	\item Hill-Climbing - score-based Bayesian network learning algorithms - using the \textit{bnlearn} R packaged (\cite{bnlearn}); and
	\item Tabu Search, a modified hill-climbing algorithm able to escape local optima, using the \textit{bnlearn} R package (\cite{bnlearn}).
\end{itemize}

\begin{equation}
{ r }_{ y|x }^{ * }=sign({ r }_{ xy })\cdot \sqrt { 1-\frac { E{ (Y-E(Y|X)) }^{ 2 } }{ var(Y) } }
\end{equation}

\noindent where $ { r }_{ xy } $ is the Pearson's correlation coefficient, \textit{var} is variance, and the expression inside the square root is a generalized measure of correlation (GMC) defined in \cite{Zheng2012}.

Finally, where possible, exploratory mediation analyses were conducted using the \textit{medmod} R package (\cite{medmod}). Sobel tests were performed to evaluate the significance of also considering the mediation effect, using the \textit{powerMediation} R package (\cite{powerMediation}).

\section{Results}

\subsection{Statistics and impact of body position}

\begin{figure}[ht]
\centering
\includegraphics[height=0.55\columnwidth]{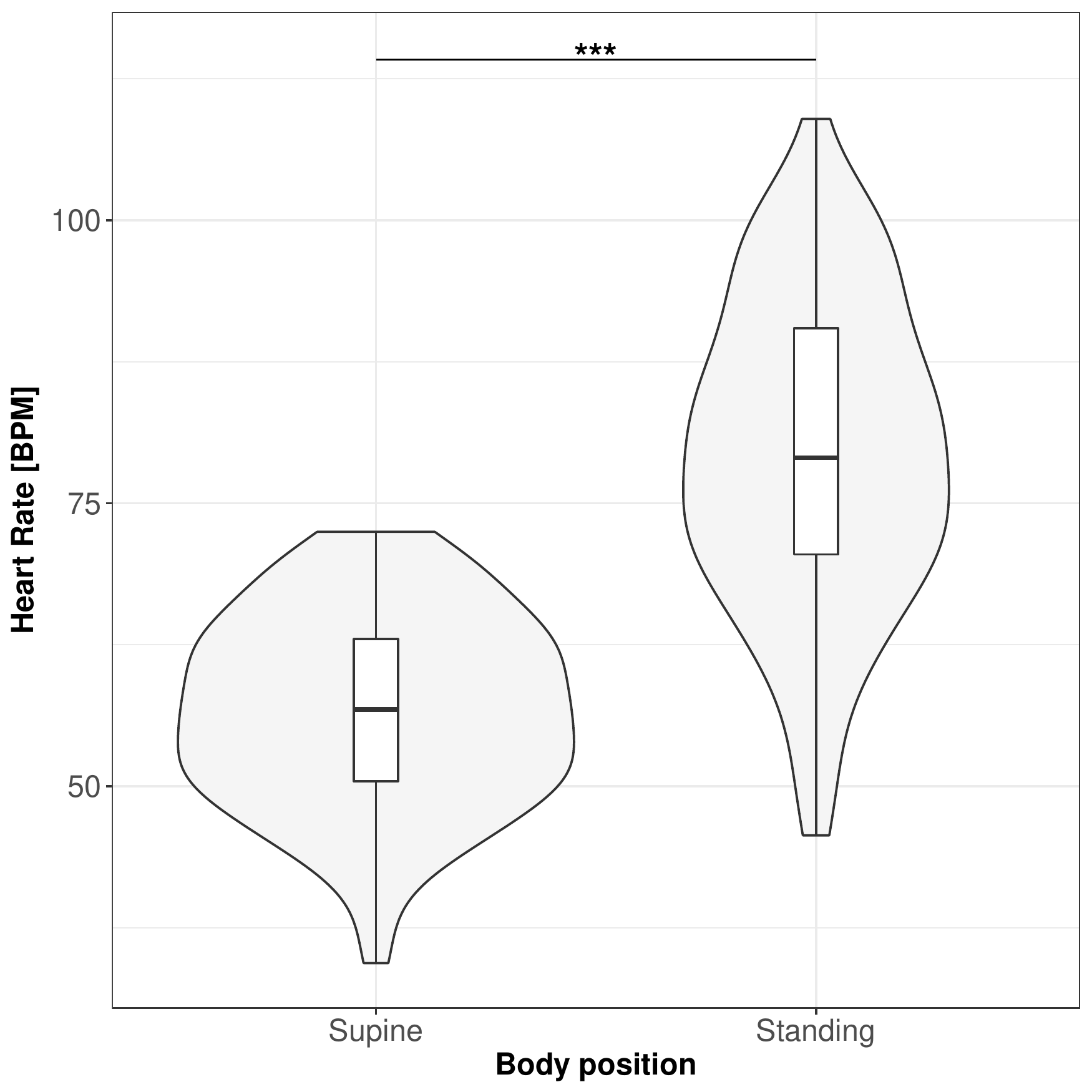}
\caption{The exploratory statistic of mean heart rate (HR) for the supine and standing body positions, along with the T paired test result.}
\label{param1}
\end{figure}

\begin{figure}[ht]
\centering
\includegraphics[height=0.55\columnwidth]{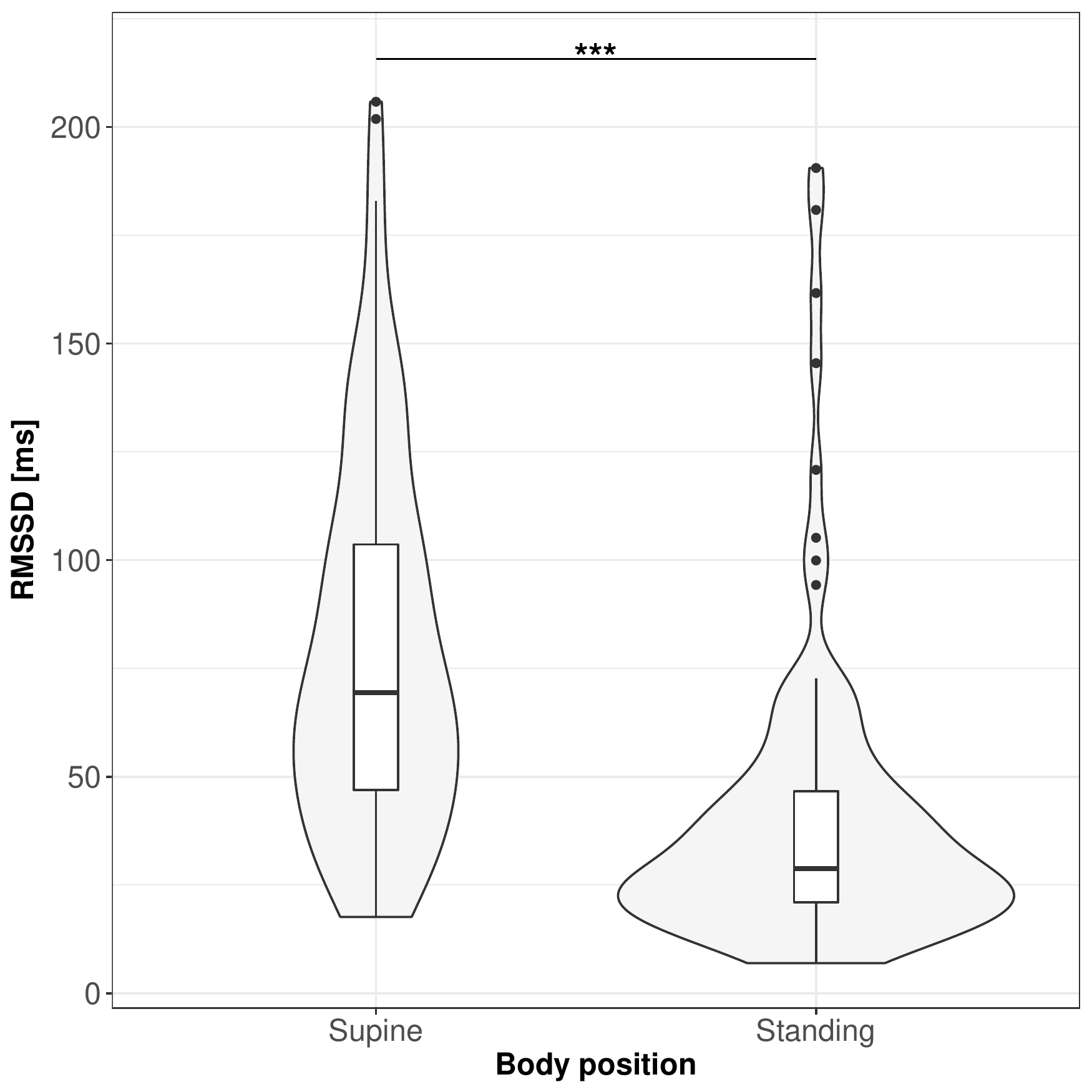}
\caption{The exploratory statistic of root-mean-square difference of successive RR intervals (RMSSD) for the supine and standing body positions, along with the T paired test result.}
\label{param2}
\end{figure}

\begin{figure}[ht]
\centering
\includegraphics[height=0.55\columnwidth]{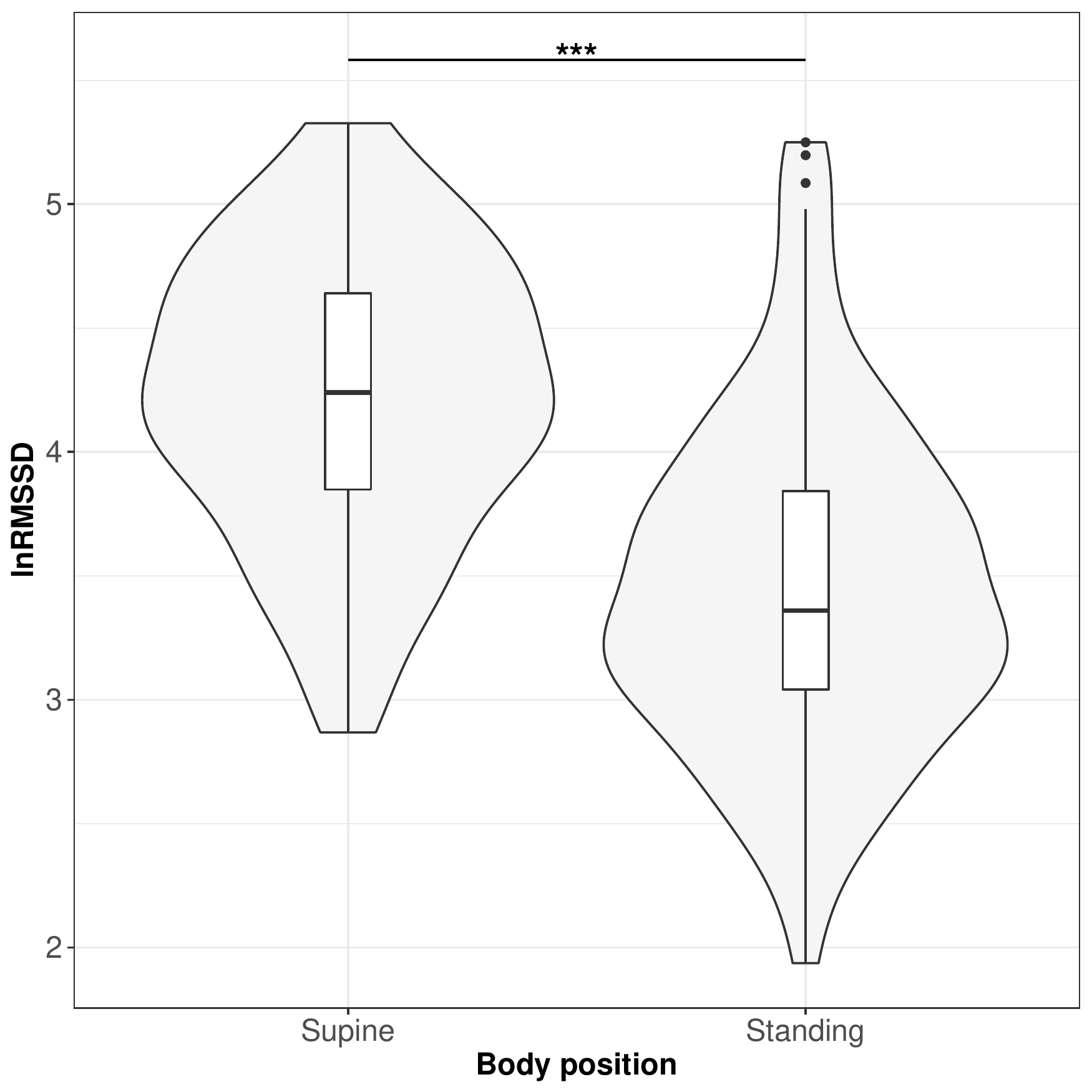}
\caption{The exploratory statistic of natural logarithm of root-mean-square difference of successive RR intervals (lnRMSSD) for the supine and standing body positions, along with the Wilcoxon rank paired test result.}
\label{param3}
\end{figure}

\begin{figure}[ht]
\centering
\includegraphics[height=0.55\columnwidth]{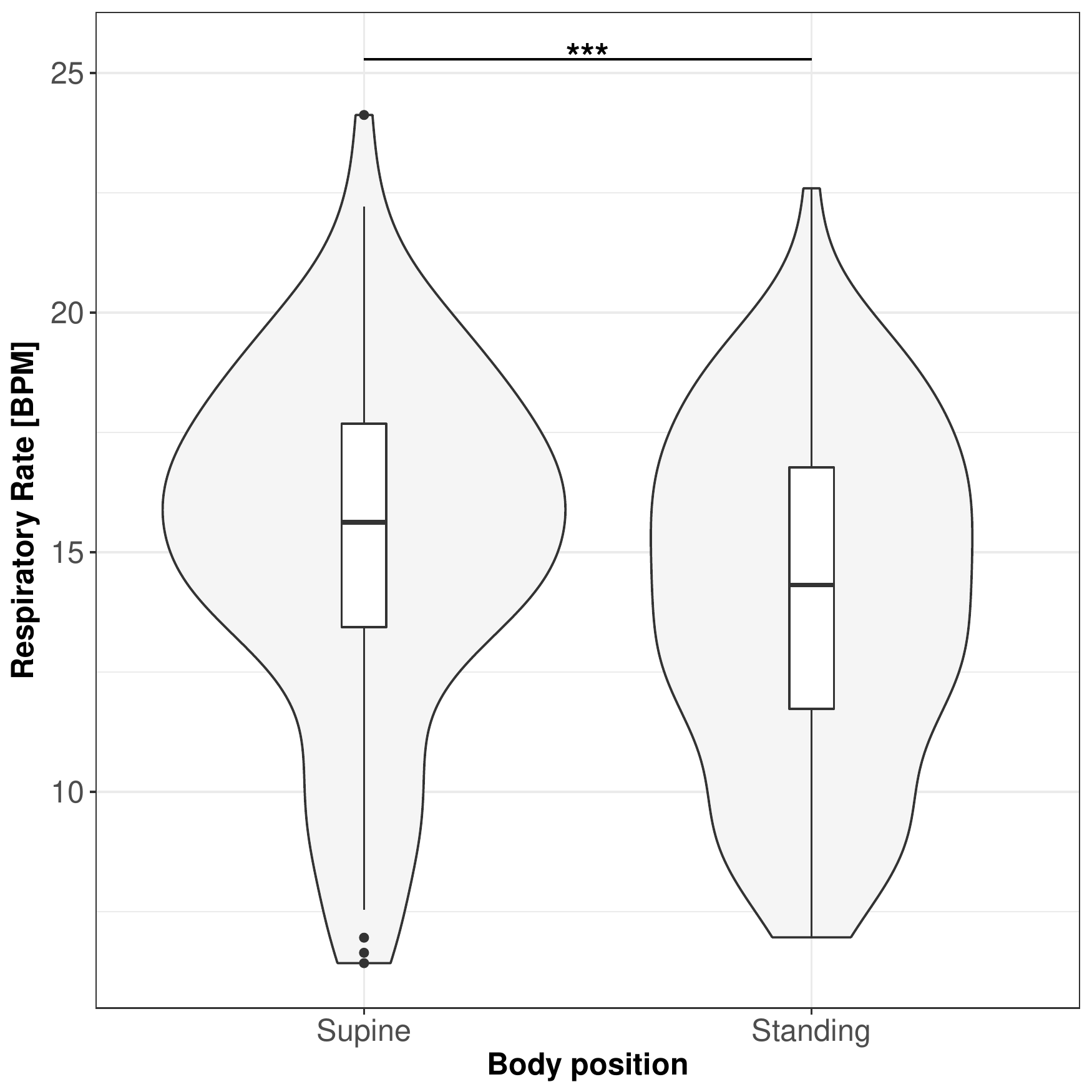}
\caption{The exploratory statistic of mean respiratory rate (RR) for the supine and standing body positions, along with the T paired test result.}
\label{param4}
\end{figure}

\begin{figure}[ht]
\centering
\includegraphics[height=0.55\columnwidth]{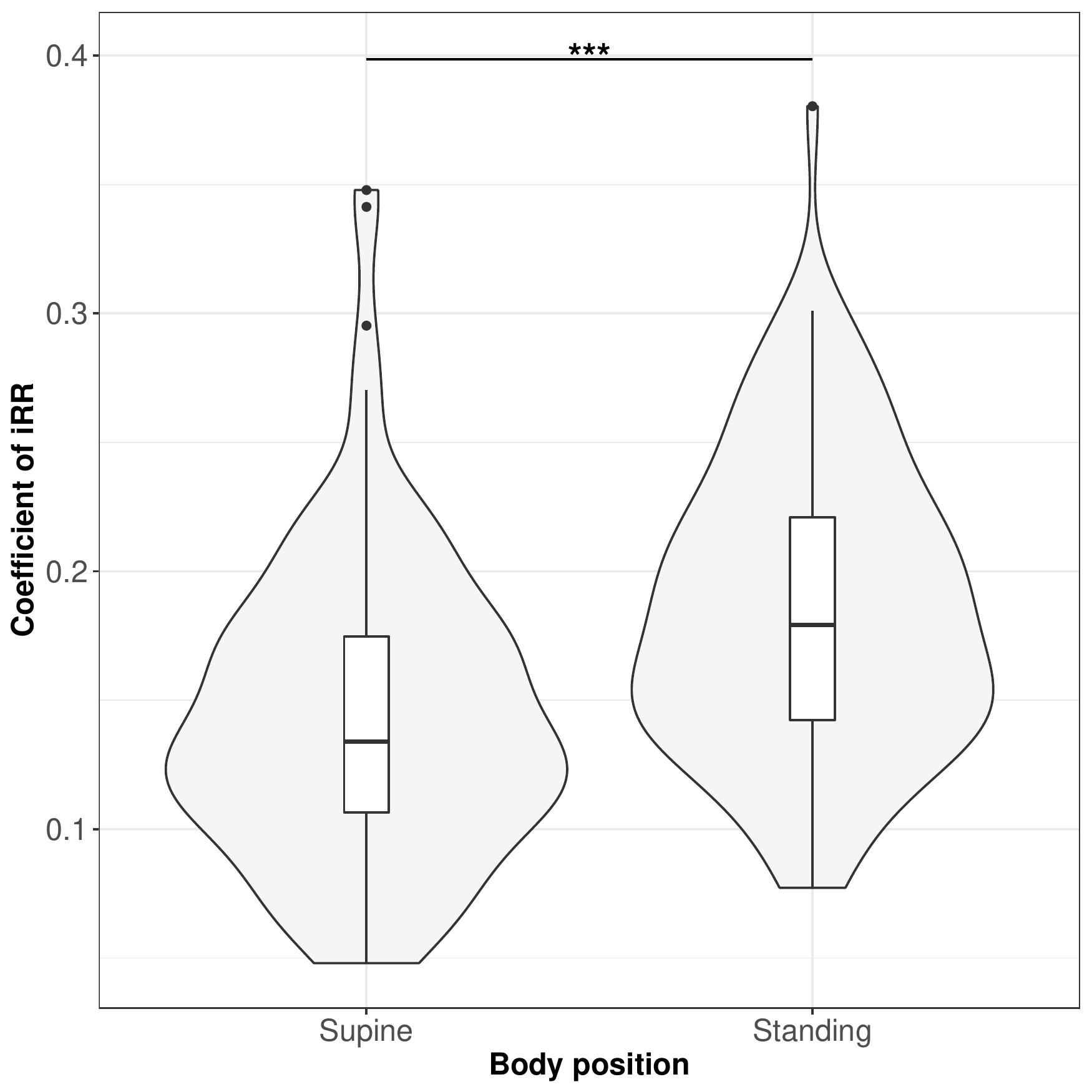}
\caption{The exploratory statistic of standard deviation of instantaneous breathing rate (calculated between inspiratory onsets - iRR), normalized to mean iRR, for the supine and standing body positions, along with the Wilcoxon rank paired test result.}
\label{param5}
\end{figure}

\begin{figure}[ht]
\centering
\includegraphics[height=0.55\columnwidth]{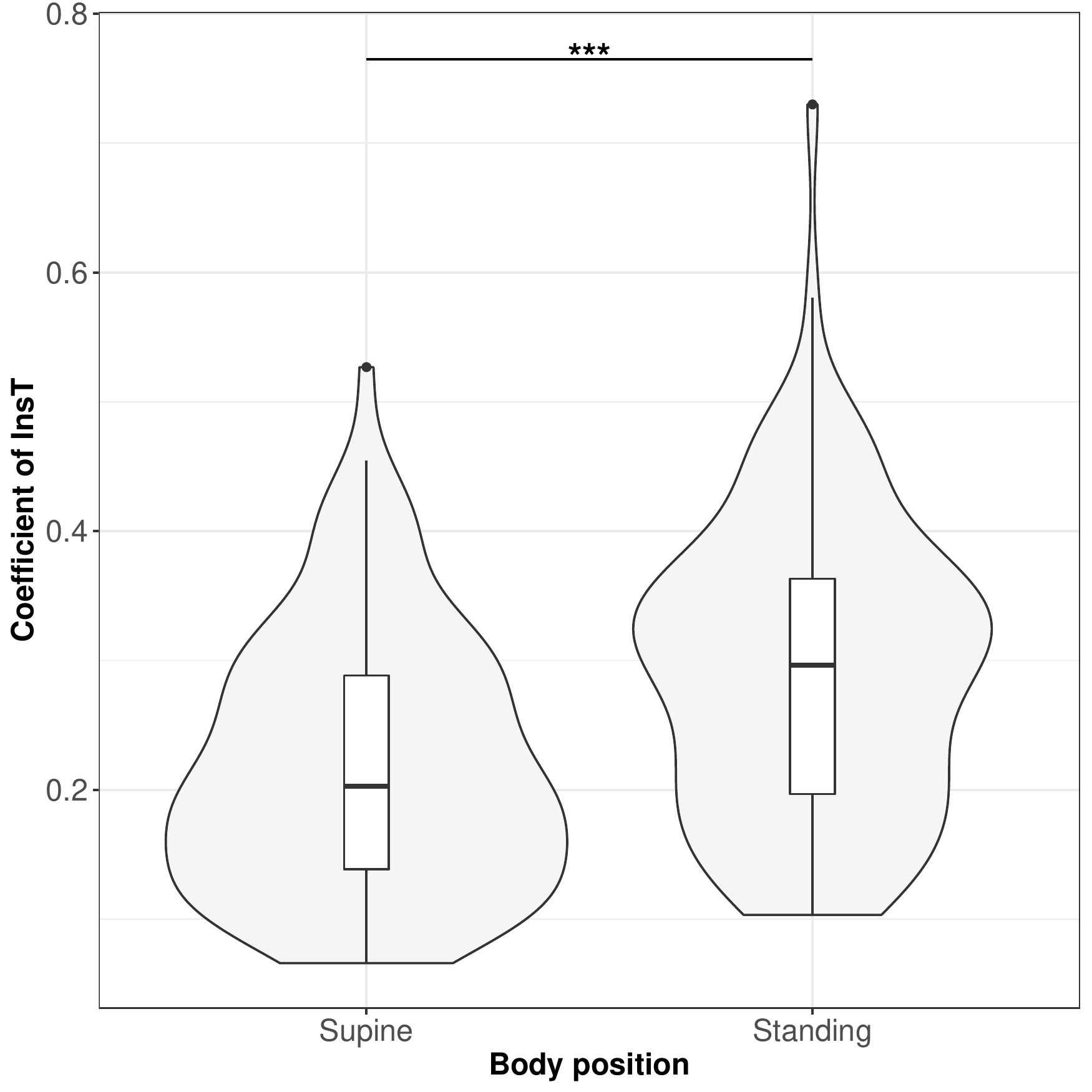}
\caption{The exploratory statistic of standard deviation of the durations of inspiratory phases (InsT), normalized to mean InsT, for the supine and standing body positions, along with the Wilcoxon rank paired test result.}
\label{param6}
\end{figure}

\begin{figure}[ht]
\centering
\includegraphics[height=0.55\columnwidth]{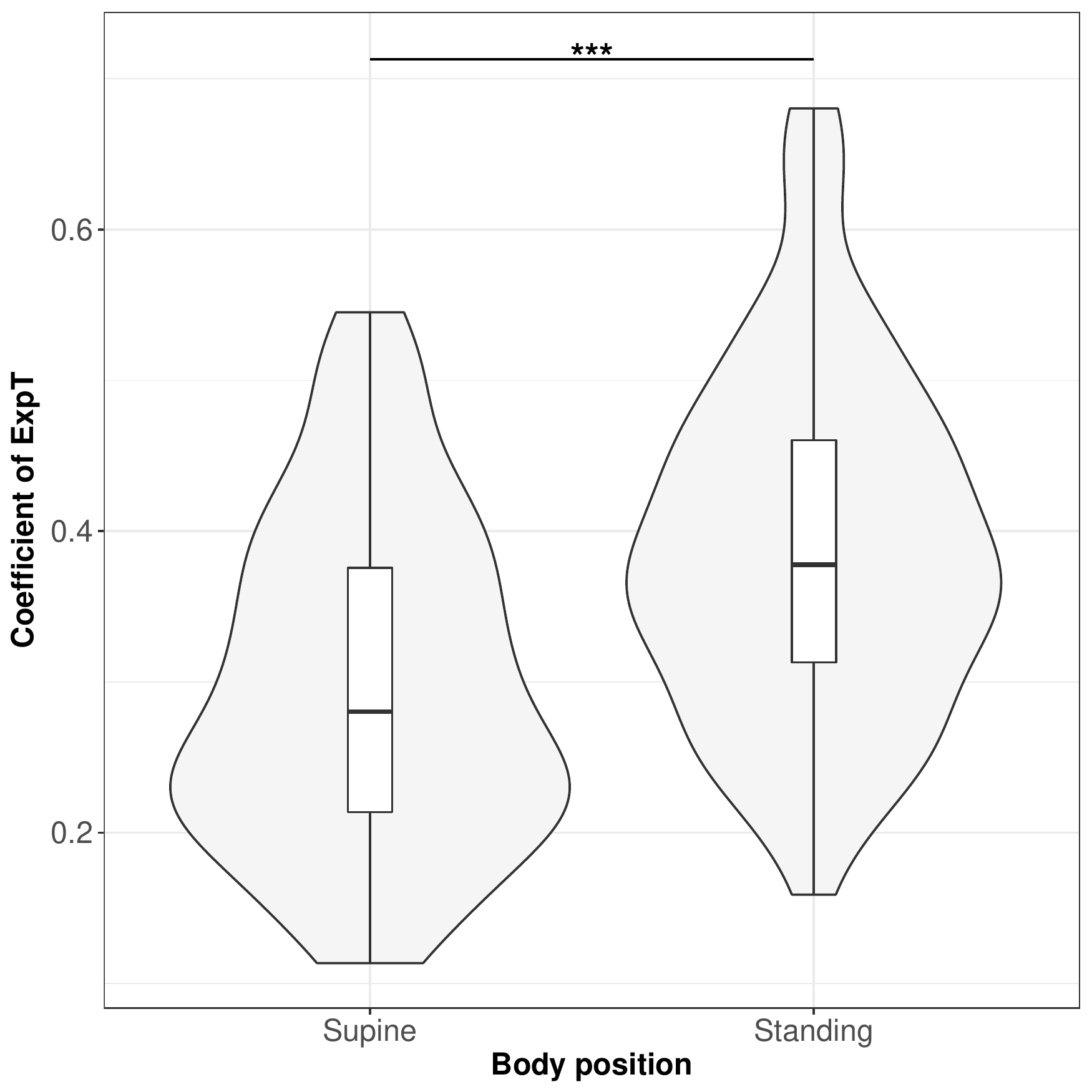}
\caption{The exploratory statistic of standard deviation of the durations of expiratory phases (ExpT), normalized to mean ExpT, for the supine and standing body positions, along with the Wilcoxon rank paired test result.}
\label{param7}
\end{figure}

\begin{figure}[ht]
\centering
\includegraphics[height=0.55\columnwidth]{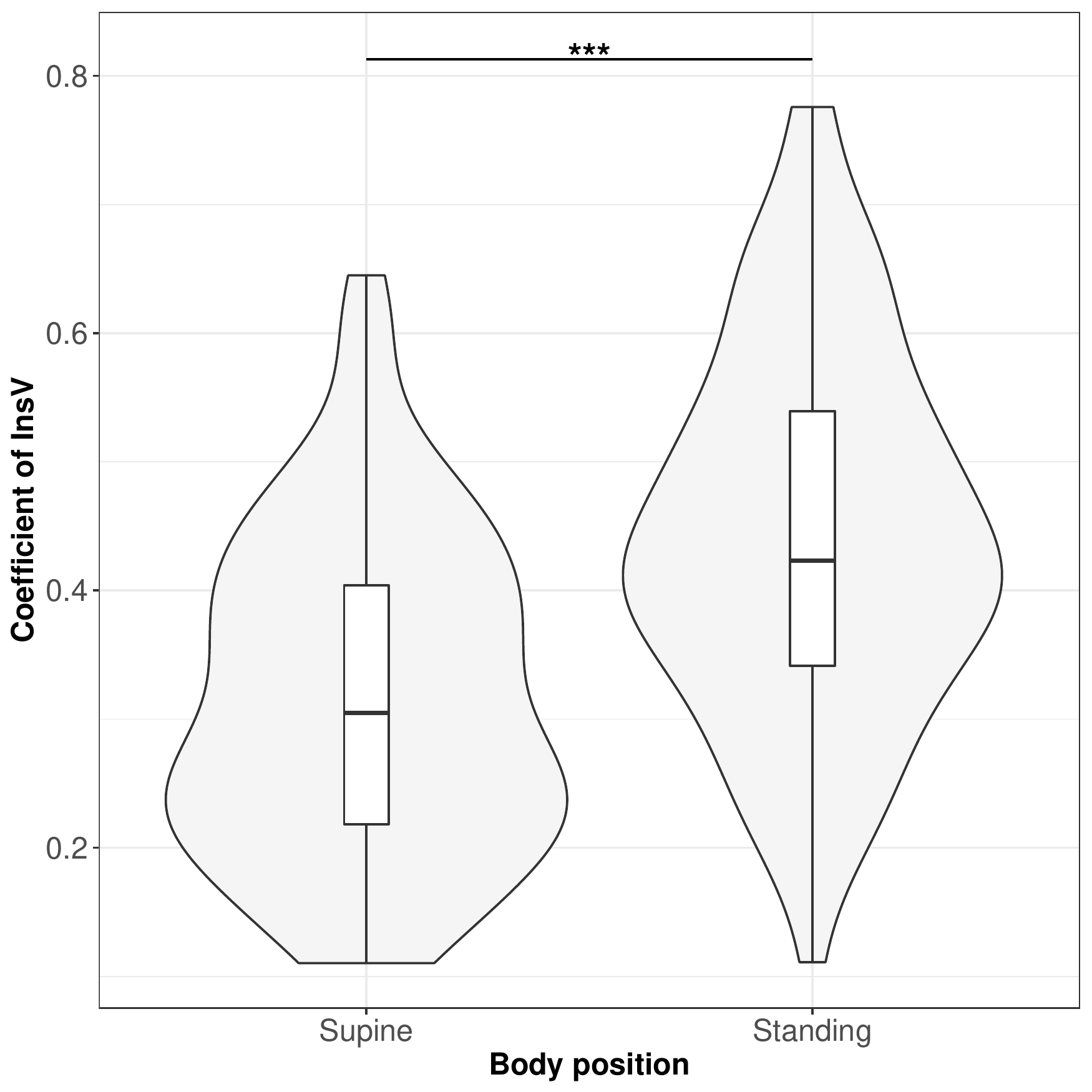}
\caption{The exploratory statistic of standard deviation of the amplitudes of expiratory phases (InsV), normalized to mean InsV, for the supine and standing body positions, along with the Wilcoxon rank paired test result.}
\label{param8}
\end{figure}

\begin{figure}[ht]
\centering
\includegraphics[height=0.55\columnwidth]{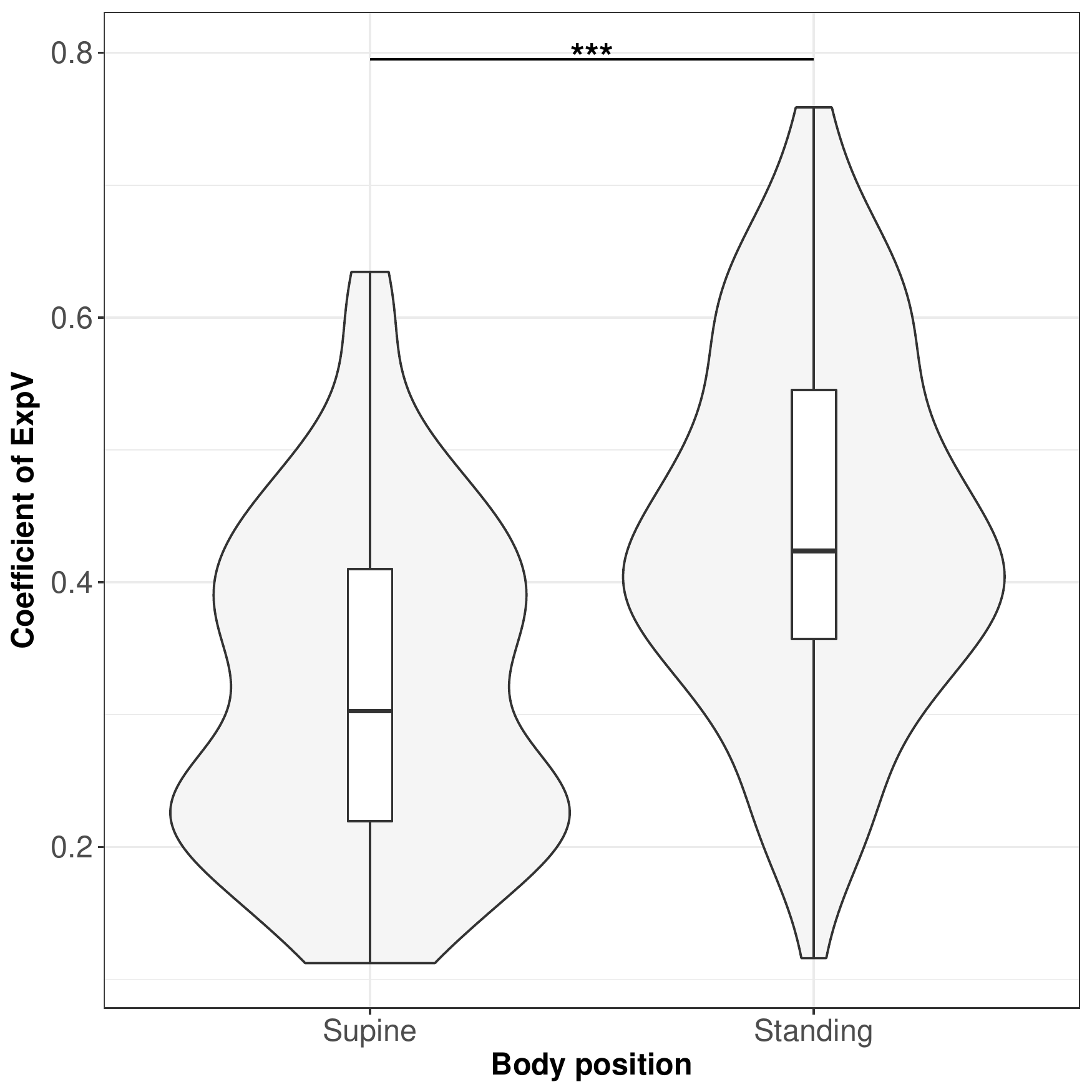}
\caption{The exploratory statistic of standard deviation of the amplitudes of expiratory phases (ExpV), normalized to mean ExpV, for the supine and standing body positions, along with the Wilcoxon rank paired test result.}
\label{param9}
\end{figure}

\begin{figure}[ht]
\centering
\includegraphics[height=0.55\columnwidth]{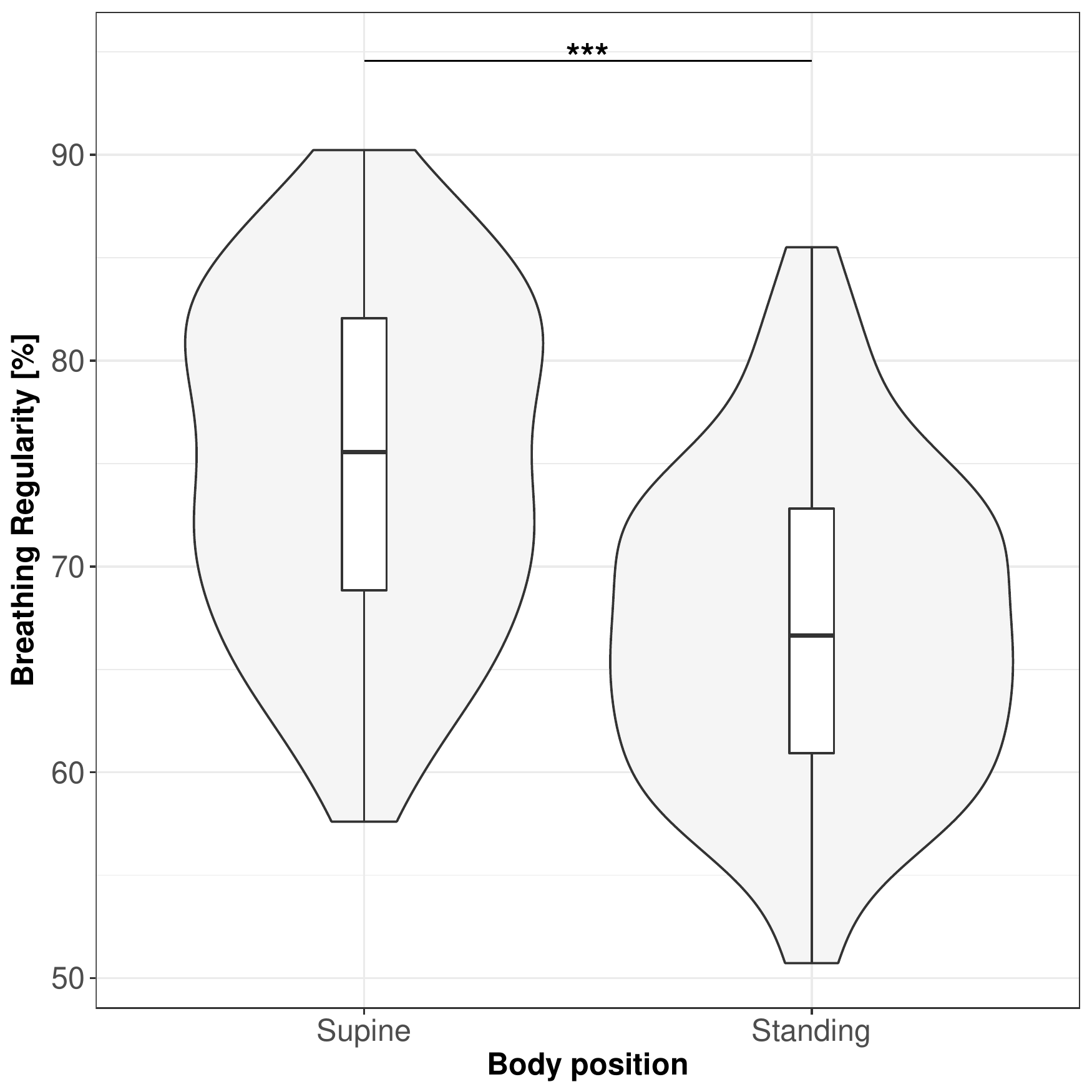}
\caption{The exploratory statistic of breathing regularity (BR) for the supine and standing body positions, along with the Wilcoxon rank paired test result.}
\label{param10}
\end{figure}

\noindent Figures \ref{param1}-\ref{param10} provide exploratory summaries (using violin and box plots) across body positions for all considered parameters, along with the paired test results. All parameters have statistically significant differences between body positions. As expected, HR was greater when standing; the reverse was true for RMSSD along with all respiratory parameters.

The significant Bayesian correlation coefficients are presented in Table \ref{bayes}. Results for supine are above the diagonal; for standing - below. Low correlations between cardiac and respiratory parameters (bottom left and top right corners of the table) suggested moderate connections between the analyzed parameters.

\subsection{Causal paths discovery and mediation analysis}

\begin{figure}[ht]
\centering
\includegraphics[height=0.55\columnwidth]{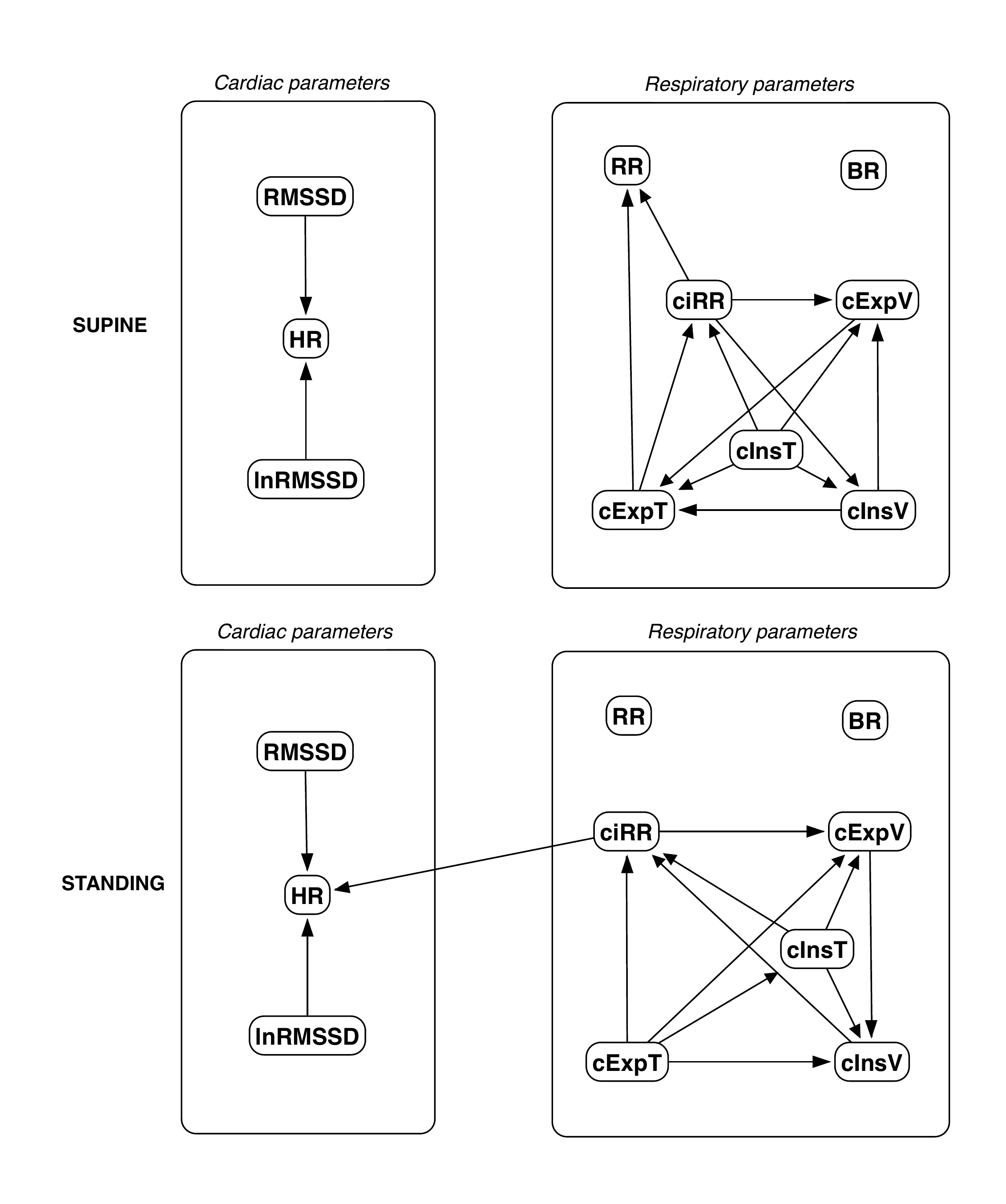}
\caption{Causal paths discovered for supine and standing body positions using generalized correlations (GC); relationships between RMSSD and lnRMSSD, and between BR and its input coefficients, are ignored.}
\label{met1}
\end{figure}

\begin{figure}[ht]
\centering
\includegraphics[height=0.55\columnwidth]{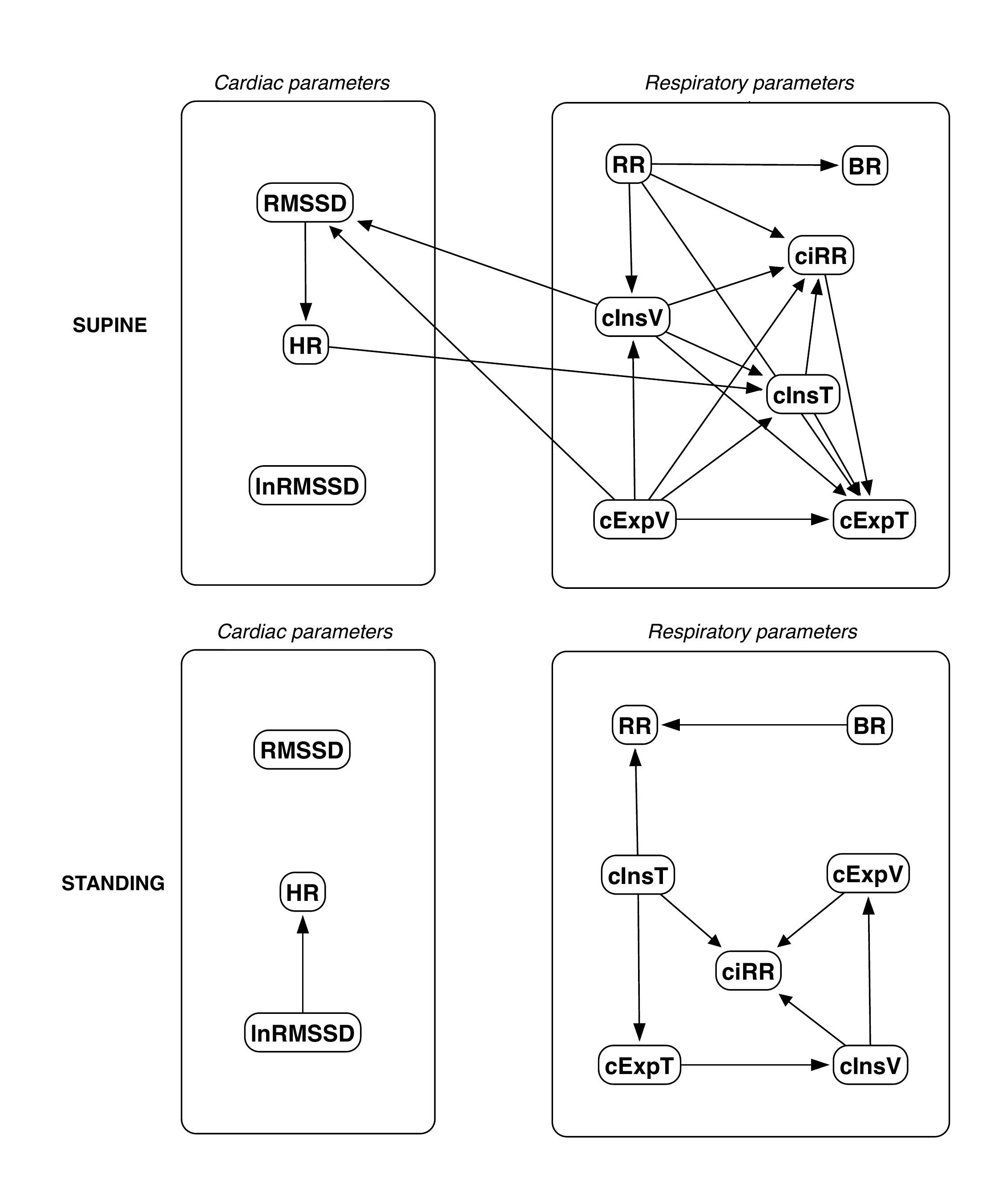}
\caption{Causal paths discovered for supine and standing body positions using causal additive modeling (CAM); relationships between RMSSD and lnRMSSD, and between BR and its input coefficients, are ignored.}
\label{met2}
\end{figure}

\begin{figure}[ht]
\centering
\includegraphics[height=0.55\columnwidth]{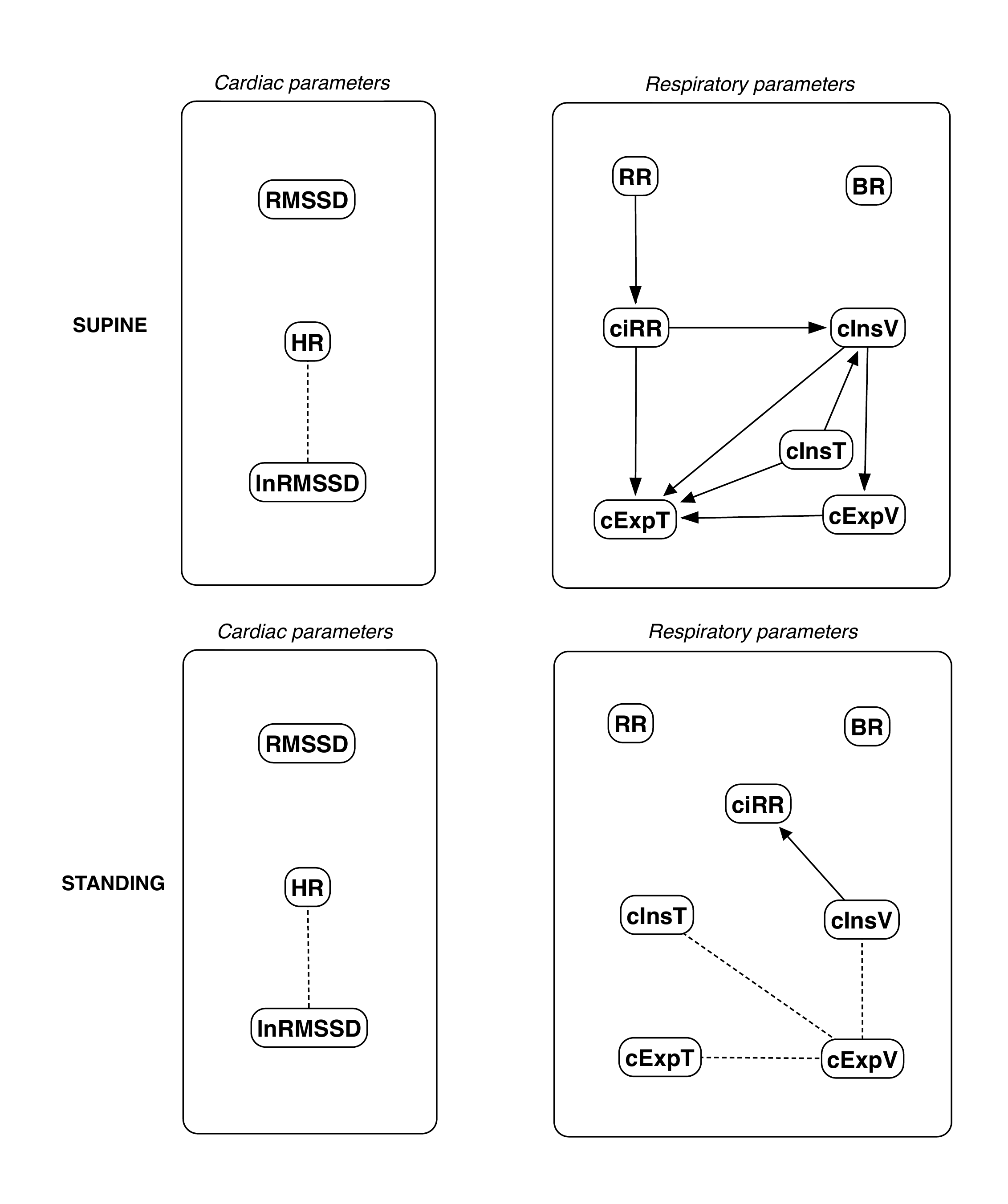}
\caption{Causal paths discovered for supine and standing body positions using fast greedy equivalence search (FGES); relationships between RMSSD and lnRMSSD, and between BR and its input coefficients, are ignored.}
\label{met3}
\end{figure}

\begin{figure}[ht]
\centering
\includegraphics[height=0.55\columnwidth]{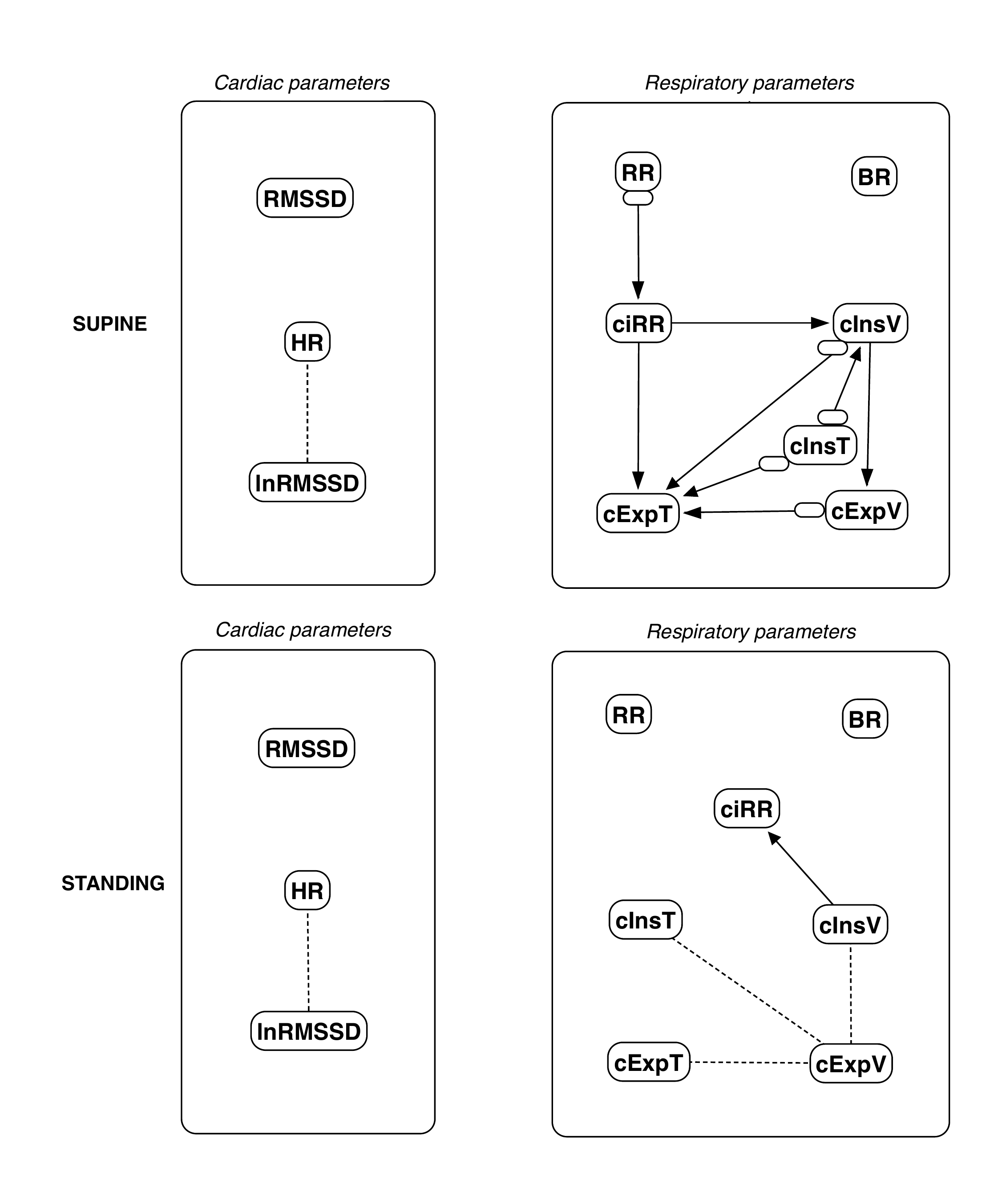}
\caption{Causal paths discovered for supine and standing body positions using greedy fast causal inference (GFCI); relationships between RMSSD and lnRMSSD, and between BR and its input coefficients, are ignored. Inclined circles at the beginnings of arrows indicate either a presented direction, an unmeasured confounder, or both.}
\label{met4}
\end{figure}

\begin{figure}[ht]
\centering
\includegraphics[height=0.55\columnwidth]{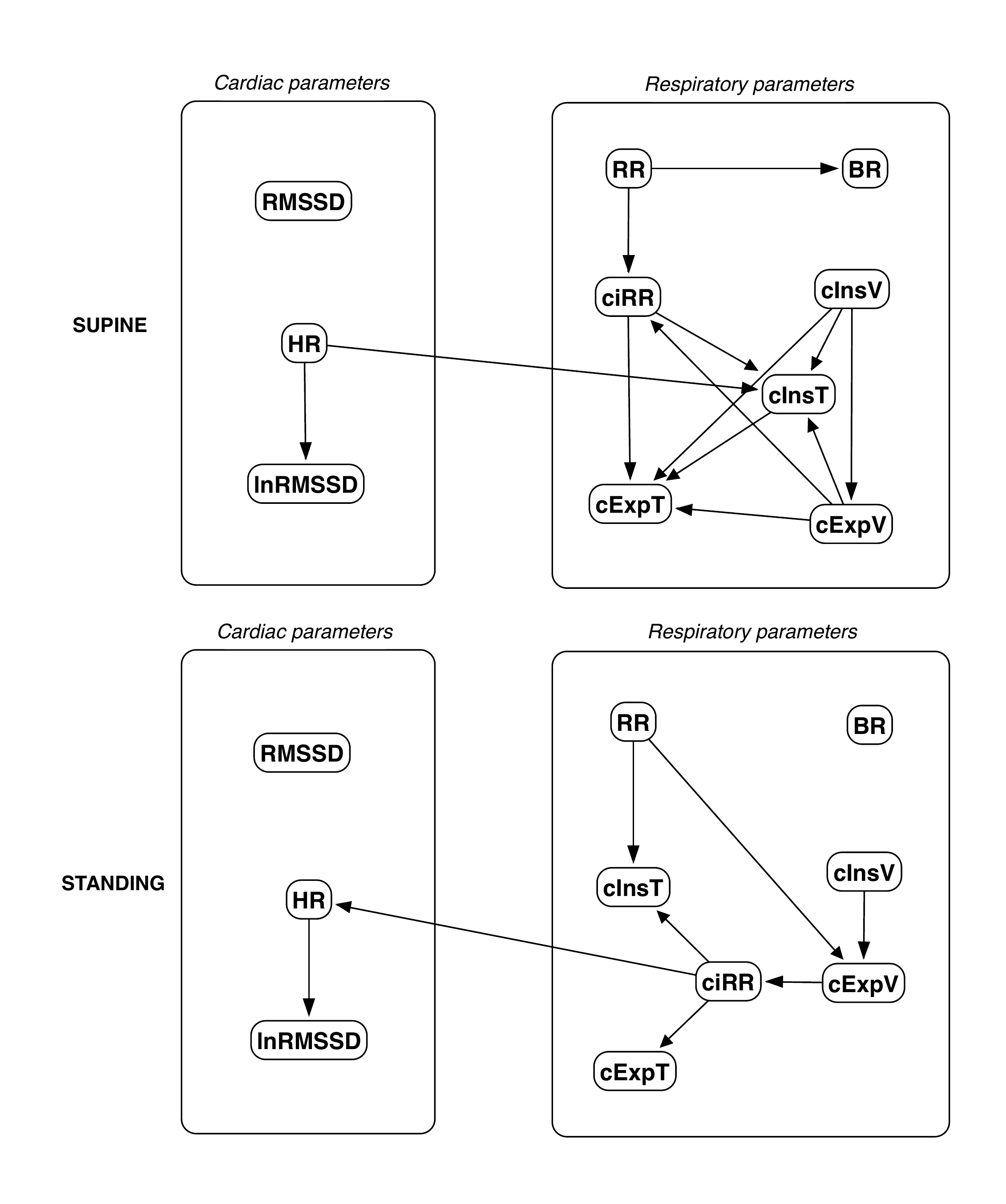}
\caption{Causal paths discovered for supine and standing body positions using the Hill Climbing Bayesian network learning algorithm (HC); relationships between RMSSD and lnRMSSD, and between BR and its input coefficients, are ignored.}
\label{met5}
\end{figure}

\begin{figure}[ht]
\centering
\includegraphics[height=0.55\columnwidth]{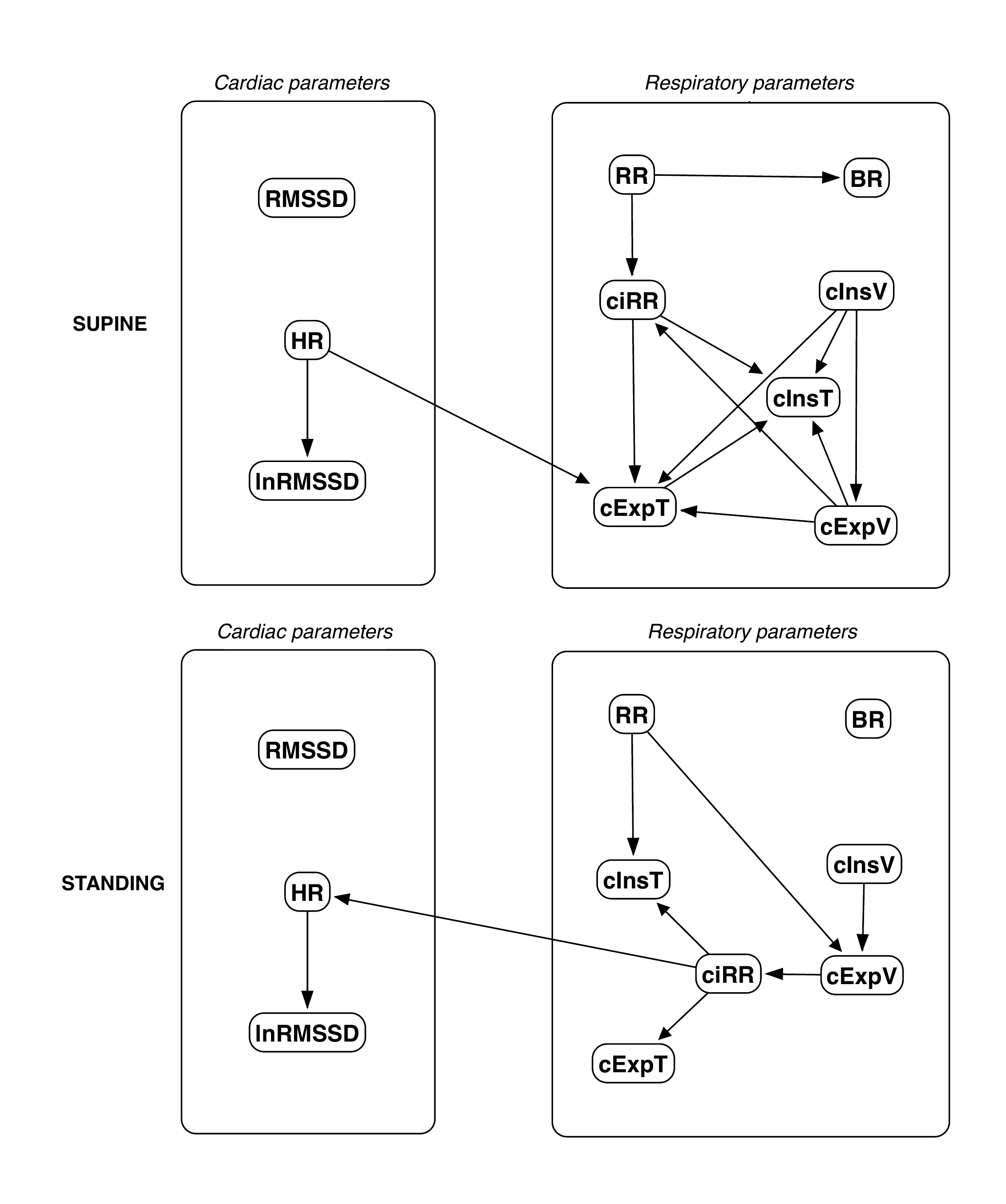}
\caption{Causal paths discovered for supine and standing body positions using Tabu Search (TS); relationships between RMSSD and lnRMSSD, and between BR and its input coefficients, are ignored.}
\label{met6}
\end{figure}

\noindent The discovered causal paths (ignoring the relationships between RMSSD and lnRMSSD, and between BR and its input coefficients) for the supine and standing positions are presented in Figures \ref{met1}-\ref{met6}, separately for each of the considered methods.

Th connections between cardiac parameters seem equivocal because GC and CAM suggested the direction from RMSSD to HR for both body positions, the greedy algorithms (FGES and GFCI) could not determine the direction, while the Bayesian network methods (HC and Tabu) recommended assuming that the right direction is the opposite.

For respiratory parameters and supine body positions, 5 of the 6 methods showed that cInsT causes cExpT, and 5 out of 6 also that cInsV causes cExpV. Furthermore, all methods indicate that cExpV causes cExpT, and that cInsV causes cExpT.

The findings for the standing body position were different. Only 3 of the 6 methods confirmed the direction from cInsV to cExpV (2 were ambiguous). InsT seems to be connected much more weakly with ExpT. cInsV and cExpV appear not to cause cExpT to the same extent as for the supine position. Several loops are present between cInsV, cExpV, and ciRR. One should note another connection, in which cInsV causes ciRR, indirectly or directly, via cExpV.

Three methods propose four connections between cardiac and respiratory parameters in supine body positions:

\begin{itemize}
	\item cInsV $\rightarrow$ RMSSD (CAM),
	\item cExpV $\rightarrow$ RMSSD (CAM),
	\item HR $\rightarrow$ cInsT (CAM and HC), and
	\item HR $\rightarrow$ cExpT (Tabu).
\end{itemize}

These all indicate the occurrence of a relatively complex connection even in the most static cases. The relationships appear to be weak. Several paths can be created (in parallel); however, we think one of them may suggest the general direction for cardiorespiratory data during supine rest:

\noindent \fbox{\small \centerline{ \textbf{Tidal Volume} $\rightarrow$ \textbf{Heart Activity Variation} $\rightarrow$ \textbf{Average Heart Activity} $\rightarrow$ \textbf{Respiratory Timing}}}

Three methods (GC, HC, and Tabu) indicate the connection ciRR $\rightarrow$ HR for the standing body position, implying the general rule for standing to be:

\noindent \fbox{\small \centerline{ \textbf{Normalized Respiratory Activity Variation} $\rightarrow$ \textbf{Average Heart Activity}} }

Finally, we set several paths to test for mediation effects:

\begin{itemize}
	\item (1) RMSSD $\rightarrow$ HR $\rightarrow$ cInsT (supine),
	\item (2) HR $\rightarrow$ cInsT $\rightarrow$ cExpT (supine),
	\item (3) HR $\rightarrow$ cInsT $\rightarrow$ cInsV (supine),
	\item (4) cInsT $\rightarrow$ ciRR $\rightarrow$ HR (standing), and
	\item (5) cInsV $\rightarrow$ ciRR $\rightarrow$ HR (standing).
\end{itemize}

The Sobel p-values for the analyzed mediations are presented in Table \ref{tab3}. While none of the considered connections have a statistically significant mediation effect, the p-values suggest tendency. It appears that the nature of these links is more non-linear, still being very light in terms of mutual correlations.

\section{Discussion}

\noindent The main finding from our analysis is that, for the supine body position and in the elite athletes group, tidal volume seems to cause heart activity variation, then the latter causes average heart activity, which appears to affect the timing of inspiratory and expiratory phases. The relations are mild and this statement is not supported by all methods, which is not to say that any oppose it. For the standing body position, the causal relations are weaker. The most important remains that in which normalized respiratory activity variation causes average heart activity. On the other hand, for these conditions, more of the cross-correlations between cardiac and respiratory parameters were statistically significant.

This suggests the need to consider activity measures from both systems; however, in the common practice, only ECG analyses are usually carried out. The simplest, but still very informative, parameters of heart activity are mean heart rate and root-mean-square difference of successive RR intervals. The first enables study of the average value of the rhythm, while the other shows its diversity.

The concept of using HRV-related data in sports analysis has been already proposed in many applications, e.g.:

\begin{itemize}
	\item quantitative assessment of training load (\cite{Saboul2016}),
	\item monitoring of weekly HRV in futsal players during the preseason to evaluate high vagal activity (\cite{Nakamura2016}),
	\item progressive sympathetic predominance at peak training load as a performance prediction factor in recreational marathon runners (\cite{Triposkiadis2009}),
	\item parasympathetic modulation elevation over a 24h recording period induced by endurance and athletic activities (\cite{Vanderlei2008}),
	\item assessment of the parasympathetic tone resulting from training (\cite{Berkoff2007}),
	\item analysis of over-training syndrome (\cite{Dong2016}), and
	\item analysis of training adaptation (\cite{Plews2013}).
\end{itemize}

However, in the presented works, the authors did not consider adding breathing activity to the analysis. Therefore, we performed the study with a device allowing recording of ventilation with minimal disruption of said activity. The Pneumonitor 2 was used to measure changes in thoracic impedance, which is related to changes in the amount of air in the lungs (\cite{Mlynczak2017}). From that information, we estimated the average respiratory rate, along with five coefficients specifying the deviation of the respiratory rate, inspiratory and expiratory phase durations and amplitudes (related to volumes), allowing creation of a novel index, breathing regularity. Consequently, we parameterized cardiac and respiratory activity with indexes, which estimate a mean value and variation.

This approach connected with findings for supine body position create an interesting cardiorespiratory loop between systems. Several studies proposed to consider multi-directionality in the coupling of the cardiovascular and respiratory systems (\cite{Porta2013,Platisa2016,Radovanovic2018}). More importantly, the relation seems to combine (in terms of a specific mathematical framework) several physiological mechanisms. The first "arrow" indirectly describes the RSA phenomenon (\cite{Shaffer2014, McCraty2015}). Some have already observed that the respiratory centers can modulate the frequency of the heart through the vagal sinus node intervention (\cite{Eckberg2009}).

The last connection appears related to cardiorespiratory coupling, described by \cite{Sobiech2017} as a relation between the histograms of R peaks appearing before the inspiratory onset and of peaks appearing just after.

The lack of a direct "arrow" from heart activity variation, namely RMSSD in this study, to breathing corresponds with the findings of \cite{Saboul2013,Sala2016}, wherein this index did not correlate with breathing, neither spontaneous nor controlled.

In that context, causality inference is a promising mathematical tool to expand the current analytical framework. Complex biological responses appear to be the right input, even when results are quite diverse.

All the techniques mentioned in the Introduction are based on pre-processed time series or beat-by-beat sequences. They allow evaluation of different time resolutions, in order to focus on the specific condition and subject. We believe that this is the right approach, and that for a better grasp of how to prepare a sport-sensitive biomarker, it should be preceded by a more general approach. The approach would be time-independent, more holistic (considering a whole group of participants), and not based on prior medical knowledge. As the causality frameworks were originally introduced to deal with interventional variables, they can be even used for prediction, e.g., when several changes in a training program can be interpreted as changes to the system. As this is a retrospective study, it is also assumed that interventions are impossible, and the networks are created with no prior knowledge. Causal search relies on passive observation.

It appears debatable whether correct causal explanations can be chosen only by looking at observed data. On the other hand prior knowledge would enable acquisition of answers to causal questions without performing interventions (still difficult to manage from physiological perspective). Here, the context is different. We believe that causal analysis and cardiorespiratory relation can be a relevant supplement to already-established techniques and play the role of a biomarker (or its part) for establishing the state of the athlete. In this introductory approach, we assumed the single parameters describe specific subjects, and the analysis collects them all together.

The proposed protocol is inspired by the orthostatic maneuver, but the analysis for each positions is independent. Moreover, we hold that, from a causal discovery perspective, one can consider the entire segment of the signal, instead of subsections like the adaptation after standing up. Looking at the entire segment makes the analysis simpler from an operator’s perspective.

As the estimated causal structures seem mild, analyzing supine-to-standing changes would make the method more robust (by considering changes in intrathoracic pressure or differences in venous return characteristics). \cite{Radovanovic2018} reported that even slight change of body position may change the direction of the relationships. \cite{Sobiech2017} also suggested that the results can be reliably analyzed only during static conditions and that the effect is the strongest in a resting state. However, in our opinion, the differences between body positions may have a significant impact on the analysis of cardiorespiratory data and should not be ignored. These differences, established for two body positions, may serve as an additional input for determination of the adaptation profile.

The effect observed for standing body position has already been studied by \cite{Sala2016}, who analyzed on a basic level the effectiveness of assessment of the balance change in the autonomic system during active verticalization. They observed that this change should be stronger in athletes and assumed that it is associated with improved physical capacity. Apart from that, increased coupling between heart period and arterial pressure in response to postural changes was reported in \cite{Silvani2017}.

\subsection{Use of causal discovery framework}

\noindent The findings presented in this study open up a novel domain of possible parameterization of physiological connections. We think the framework can be used to profile athletes, analyzing trends throughout the training schedule and testing changes in relations' direction and strength relative to improved adaptation, etc.

Causal inference, even if used to search for graphs between analyzed parameters, was originally proposed to evaluate the effects of interventions, dividing variables into interventional and observational sets. In both sports medicine and daily practice, modification of the training can be regarded as such an intervention.

The paths discovered by the various algorithms should be reconciled with medical background knowledge and, even more importantly, should be verified further in a prospective study. With so many possible interventions, the paths may reduce the complexity of test protocols: one might expect to influence heart rate and its variability by changing the depth, not the frequency, of breathing.

\subsection{Limitations of the study}

\noindent There were only 116 participants, of whom 100 were ultimately considered. The study group appears to be heterogeneous, apart from the fact that the studies were conducted in the "hot period" 3-4 months before the Olympic Games, which may suggest a state of over-training (neither questionnaires nor objective data about it were collected). Also, the device was new for all subjects. These factors might influence the results, particularly the relatively high supine respiratory rates and standing heart rate. The existence of differences between sports would also affect the conclusions. The collection of only one observation per athlete precludes reproducibility analysis.

There were no control variables assumed in the protocol. One may consider a respiratory activity to be controlled during the measurement. However, the primary objective of the registration was to measure cardiorespiratory data without any restrictions to the respiratory activity. As this is a retrospective study, we could not affect it during the analysis. This may be one of the reasons why the studied relationships were so mild.

Measurements were carried out only once for each subject, and only at rest, not in a natural environment. It is necessary to perform comparative registrations on a unified group of athletes under laboratory conditions, during the preparatory period and just before the season, in combination with a psychometric questionnaire. The results of a study in which the registration could be performed outside the laboratory, during normal training, or even with 24h Holter-based tracking of natural functioning, would yield more condensed findings.

Notably, the methods used to discover causal paths sense different aspects (the algorithms having different starting points) and can produce more liberal or conservative results. As only the "significant" pairs of causal-effect links were presented, possible bias cannot be evaluated directly. As the relevant analysis could not be performed for a few participants, we decided not to divide participants by sex or sport type.

\subsection{Further considerations}

\noindent In the presented work, we use a data-driven and time-independent approach (on a large inter-individual scale). It is considered as a starting point. As stated in the introduction, the presented approach seems to be an overview, to be performed before time-dependent methods which may allow the separate assessment of a single person, with finer time resolution. Therefore, as a next step, we plan to discover various non-linear parameters and techniques to confirm, expand and enhance the findings presented in this paper. Still, the main outcomes are coherent with the knowledge-based predictions and results reported, e.g., in \cite{Sobiech2017}.

Eckberg showed that the RSA phenomenon is not only the effect of respiration on RR intervals (or heart rate in general) but might be also treated as the response of heart rate to the respiratory modulations resulting from arterial pressure changes mediated by baroreflex (\cite{Eckberg2009}). \cite{Sobiech2017} suggested that arterial blood pressure is probably the driver (cause) of both cardiac and respiratory function. This requires further research with more modalities included in the analysis (\cite{Zhang2016}).

Other important questions include: Are the cardiorespiratory connections (direction and strength) the same in different sports, or specific to each? What other factors, like sex, height, or body mass, confound the results? These questions are especially important not only for elite athletes, but also for subjects with abnormalities, and will be studied (\cite{Sharma2017}).

\section{Conclusions}

\noindent Adding respiratory data to cardiac signals in sports medicine would provide better monitoring and evaluation of athletic training.

In the presented paper, we pose the hypothesis that average or diversity cardiac and respiratory parameters, describing overall characteristics of RR intervals and tidal volume curves, without consideration of the time resolution, are linked by causal paths. Furthermore, such an approach can precede more detailed, individual, time-series-related analysis.

Based on the data gathered from 100 elite athletes at rest in two body positions, the applied causal discovery frameworks suggested moderate connections. For supine, the general path led from tidal volume, through heart activity variation and average heart activity, to respiratory timing. For standing - from normalized respiratory activity variation to average heart activity. Different graphical structures and directions were observed for the two body positions, which may improve the resolution of the findings.

We think posterior-style causal inference and characterization may develop descriptions of cardiorespiratory connections and possibly distinguish between various groups of athletes. The method can be used to profile athletes, elaborate on modifications of their training schedules, and find objective ways to improve their competitive performance.

\section*{Conflict of Interest Statement}

The authors declare that the research was conducted in the absence of any commercial or financial relationships that could be construed as a potential conflict of interest.


\section*{Acknowledgments}

The authors thank Martin Berka for linguistic adjustments.

\noindent Marcel Młyńczak gratefully acknowledges the Institute of Metrology and Biomedical Engineering of the Faculty of Mechatronics, Warsaw University of Technology, for supporting this work.

\section*{Supplemental Data}

1. The data file with cardiac and respiratory parameters calculated from all participants for the two body positions.

\noindent 2. The R script comprising all algorithms mentioned in the text, from loading of the above data to delivery of final outcomes.

\section*{Author Contributions}

\noindent\textbf{Conceptualization:} MM HK. \\
\noindent\textbf{Data curation:} MM. \\
\noindent\textbf{Formal analysis:} MM. \\
\noindent\textbf{Investigation:} MM HK. \\
\noindent\textbf{Methodology:} MM HK. \\
\noindent\textbf{Project administration:} MM HK. \\
\noindent\textbf{Resources:} HK. \\
\noindent\textbf{Validation:} MM HK. \\
\noindent\textbf{Visualization:} MM. \\
\noindent\textbf{Writing and reviewing} MM HK. \\

\bibliographystyle{frontiersinSCNS_ENG_HUMS}

\pagebreak

\section*{Tables and Figures}

\begin{table}[h]
\centering
\caption{The information of the set of participants evaluated after excluding those with too much signal distortion; despite the lack of distinction in the paper, the table is divided into types and groups of sports, for better insight; the sport types are defined according to (\cite{Mitchell2005}), where numbers refer to the static component of heart activity expressed as \% of its maximal voluntary contraction (MVC) - low (I), medium (II), and High (III) - and letters to the dynamic component (e.g., \% of $VO_{2}max$).}
\vspace{0.2cm}
\label{tab1}
\begin{tabularx}{\columnwidth}{XX|rr|rrr|rrr}
\hline
\begin{tabular}[c]{@{}l@{}}\textbf{Group}\end{tabular} & \begin{tabular}[c]{@{}l@{}}\textbf{Sport type}\end{tabular} & \multicolumn{2}{c|}{\textbf{N}} & \multicolumn{3}{c|}{\textbf{Height}} & \multicolumn{3}{c}{\textbf{Body mass}} \\
& & \textbf{Female} & \textbf{Male} & \textbf{Min}    & \textbf{Mean} & \textbf{Max} & \textbf{Min}    & \textbf{Mean} & \textbf{Max} \\ \hline
\textbf{B} & IB  &  4 & 21 & 61.0 & 82.6 & 104.2 & 170 & 193.2 & 208  \\
& IIB  &  7 &  2 & 55.0 & 64.8 &  97.7 & 167 & 174.7 & 193  \\
& IIIB  &  4 &  8 & 53.2 & 79.7 & 151.0 & 158 & 174.3 & 197  \\
\textbf{C} & IC  &  1 &  4 & 55.1 & 71.7 &  85.2 & 169 & 176.0 & 190  \\
& IIC  & 12 & 25 & 49.1 & 80.7 & 115.0 & 162 & 185.2 & 207  \\
& IIIC  &  4 &  8 & 62.7 & 75.2 &  87.7 & 171 & 179.5 & 189  \\ \hline
\end{tabularx}
\end{table}

\begin{table}[h]
\centering
\caption{Bayesian correlation coefficients calculated between parameters and presented when significant. Results for supine are above the diagonal; for standing - below.}
\vspace{0.2cm}
\label{bayes}
\begin{tabular}{l|rrr|rrrrrrr}
\hline
                 & \textbf{HR} & \textbf{RMSSD} & \textbf{lnRMSSD} & \textbf{RR} & \textbf{ciRR} & \textbf{cInsT} & \textbf{cExpT} & \textbf{cInsV} & \textbf{cExpV} & \textbf{BR} \\ \hline
\textbf{HR}      & -           & -0.36          & -0.41            &             &               & -0.17          &                &                &                &             \\
\textbf{RMSSD}   & -0.47       & -              & 0.95             &             &               &                &                &                &                &             \\
\textbf{lnRMSSD} & -0.51       & 0.91           & -                &             &               &                &                &                &                &             \\ \hline
\textbf{RR}      &             & 0.14           &                  & -           & -0.42         & -0.14          & -0.22          &                &                & 0.16        \\
\textbf{ciRR}    & 0.22        & -0.15          & -0.17            & -0.19       & -             & 0.69           & 0.73           & 0.62           & 0.62           & -0.81       \\
\textbf{cInsT}   & 0.16        & -0.17          & -0.18            &             & 0.59          & -              & 0.70           & 0.63           & 0.63           & -0.84       \\
\textbf{cExpT}   & 0.13        &                & -0.16            &             & 0.66          & 0.39           & -              & 0.62           & 0.61           & -0.84       \\
\textbf{cInsV}   &             &                &                  &             & 0.51          & 0.36           & 0.44           & -              & 0.96           & -0.91       \\
\textbf{cExpV}   &             &                &                  &             & 0.56          & 0.40           & 0.48           & 0.90           & -              & -0.91       \\
\textbf{BR}      & -0.16       &                & 0.15             &             & -0.79         & -0.68          & -0.73          & -0.85          & -0.88          & -          \\ \hline
\end{tabular}
\end{table}

\begin{table}[h]
\centering
\caption{The Sobel's p-values assessing the significance of the mediation effect for the chosen causal path connections discovered for the supine and standing body positions.}
\vspace{0.2cm}
\label{tab3}
\begin{tabularx}{\columnwidth}{X|l|r}
\hline
\textbf{Path} & \textbf{Position} & Sobel's p-value \\ \hline
RMSSD $\rightarrow$ HR $\rightarrow$ cInsT & Supine & 0.073  \\
HR $\rightarrow$ cInsT $\rightarrow$ cExpT & $-||-$ & 0.086 \\
HR $\rightarrow$ cInsT $\rightarrow$ cInsV & $-||-$ & 0.088 \\
cInsT $\rightarrow$ ciRR $\rightarrow$ HR & Standing & 0.105 \\
cInsV $\rightarrow$ ciRR $\rightarrow$ HR & $-||-$ & 0.058 \\
\hline
\end{tabularx}
\end{table}


\end{document}